\documentclass[structabstract]{aa}  
\usepackage{graphicx}
\usepackage[varg]{txfonts}
\usepackage[english]{babel}
\begin{document}

\title{Low Delta-V near-Earth asteroids: \\ A survey of suitable targets for space missions*}
\thanks{Based on observations carried out at the European Southern 
Observatory - NTT (Programs 075.C-0282, 076.C-0577 and 078.C-0141), NASA - IRTF (Program 087) and Telescopio Nazionale Galileo (Sessions AOT14-TAC41, AOT15-TAC69 and AOT26-TAC32.)}

   \subtitle{}

   \author{ S. Ieva \inst{1,2,3},
          E. Dotto \inst{1},
          D. Perna \inst{2},
           M. A. Barucci \inst{2},
          F. Bernardi \inst{4},
        S. Fornasier \inst{2,5},
          F. De Luise \inst{6},
         E. Perozzi  \inst{7,8,9},
          A. Rossi \inst{10},
         J.R. Brucato \inst{11}
}

\offprints{S. Ieva}

\institute{ INAF -  Osservatorio Astronomico di Roma, Via Frascati 33, 00040 Monte Porzio Catone, Rome, Italy\\
        \email{simone.ieva@oa-roma.inaf.it}
\and LESIA - Observatoire de Paris, CNRS/UPMC (Paris 6)/Univ. Paris Diderot (Paris 7), 5 place J. Janssen, 92195 Meudon, France
\and Universit\`a di Roma Tor Vergata, Via della Ricerca Scientifica 1, 00133 Rome, Italy
\and SpaceDyS, Via Mario Giuntini 63, 56023 Cascina, Pisa, Italy
\and Universit\'{e} Paris Diderot -- Sorbonne Paris Cit\'e, 4 rue Elsa Morante, 75205 Paris,  France
\and INAF - Osservatorio Astronomico di Teramo, Via Mentore Maggini snd, 64100 Teramo, Italy
\and Deimos Space, Ronda de Poniente 19, 28760 Tres Cantos,  Madrid, Spain
\and INAF - IAPS, Via Fosso del Cavaliere 100, 00133 Rome, Italy
\and ESA - NEOCC, ESRIN, Via Galileo Galilei 64, 00044 Frascati, Rome, Italy
\and IFAC- CNR, Via Madonna del Piano 10, 50019 Sesto Fiorentino, Firenze, Italy
\and INAF - Osservatorio Astrofisico di Arcetri, Largo E. Fermi 5, I-50 125 Firenze, Italy
}

   \date{Received: 11 December 2013 / Accepted: 28 June 2014  }


\textbf{}

\abstract
 {In the past decades near-Earth objects (NEOs) have become very important targets to study, since they can give us clues to the formation, evolution, and composition of the solar system. In addition, they may represent either a threat to humankind or a repository of extraterrestrial resources for suitable space-borne missions. Within this framework, the choice of next-generation mission targets and the characterisation of a potential threat to our planet deserve special attention.}
 {To date, only a small part of the 11~000 discovered NEOs have been physically characterised. From ground- and space-based observations, one can determine some basic physical properties of these objects using visible and infrared spectroscopy.}
 {We present data for 13 objects observed with different telescopes around the world (NASA-IRTF, ESO-NTT, TNG) in the  0.4 - 2.5 $\mu m $ spectral range, within the NEOSURFACE survey. Objects are chosen from among the more accessible for a rendez-vous mission. All of them are characterised by a  delta-V (the change in velocity needed for transferring a spacecraft from low-Earth orbit to a rendez-vous with NEOs) lower than 10.5 km/s, well below the solar system escape velocity (12.3 km/s).}
 {We taxonomically classify 9 of these objects for the first time. Eleven objects belong to the S-complex taxonomy, and the other 2 belong to the C-complex. We constrain the surface composition of these objects by comparing their spectra with meteorites from the RELAB database. We also compute olivine and pyroxene mineralogy for asteroids with clear evidence of pyroxene bands. Mineralogy confirms the similarity with the already found H, L, or LL ordinary chondrite analogues.}
 {}


\titlerunning{Low Delta-V Near-Earth Objects}
\authorrunning{S. Ieva et al.}

\maketitle
        %

\section{Introduction}

Near-earth objects (NEOs) are the subject of great interest in the scientific community, for both their potential usefulness for future space-borne activities and their importance for the formation and the evolution of the solar system. The current standard model for the solar system's formation predicts that the population of small bodies is the remnant of the process that led to the formation of the planets some 4.6
billion years ago. They are also invoked to explain the actual abundance on inner planets of water and organics. It has also been implied that NEOs could have played an important role in the development of life (Maurette, 2005).

Besides the delivery of water and organics, the investigation of near Earth asteroids (NEAs) and comets is particularly important because they represent a well-known threat to life on our planet. Asteroid impacts, which billions of years ago  are supposed to have delivered organics on Earth, pose a significant threat to the whole of human civilization today. Space agencies are nowadays studying mitigation techniques. Most of them, like the kinetic impactor, depend strongly on the composition of the target asteroid.

A high degree of diversity in terms of composition and spectral properties is present amongst the NEA population, with compositions ranging from siliceous to carbonaceous, from basaltic to metallic, necessarily implying completely different internal densities and strengths. The physical properties of more than 85\% of the 11~000 NEAs  discovered to date are still unknown. The spectroscopy of the asteroid's surface is correlated with the internal composition, hence structure. As such, further spectroscopic observations of NEAs would assist in the characterisation and thus suitability of future space missions. Missions to NEOs have been proven to be both technically feasible and financially viable on a low budget, making them particularly interesting for both national space agencies and private companies (Zimmer \& Messerschmid, 2011).

The accessibility from Earth of potential targets of space missions is studied by classical orbital transfer algorithms, like the Hohmann transfer formulation (Prussing, 1992), which gives the minimum energy transfer trajectory between two orbits in space. This algorithm has been used to retrieve delta-V ($\Delta V$), the velocity change needed to be applied to the spacecraft
to realise a rendez-vous mission to NEAs (Shoemaker \& Helin 1978, Perozzi et al. 2001). NEAs could be an easier target than the  Moon or a much more complicated one than Jupiter (Perozzi et al. 2010). In addition, objects with a $\Delta V$ lower than the solar system escape velocity  (12.3 km/s) are easier to reach for a manned or a robotic exploration.

In this paper we present the result of spectroscopic observations in the visible (0.4 - 0.8 $\mu m$) and near infrared ranges  (0.9 - 2.4 $\mu m $) for 13 NEAs with $\Delta V$ lower than 10.5 km/s, as computed by Lance Benner's list\footnote{http://echo.jpl.nasa.gov/$\sim$lance/delta\textunderscore{v}/delta\textunderscore{v}.rendezvous.html} of $\Delta V$.
These observations were performed  in the framework of the NEO-SURFACE survey\footnote{http://www.oa-roma.inaf.it/planet/NEOSurface.html}, which is a survey dedicated to the investigation of low $\Delta V$ and potentially hazardous asteroids (PHAs).

\section{Observations and data reduction}
The data presented in this article have been collected with various instruments working in the visible and near-infrared range. They were obtained with the New Technology Telescope (ESO-NTT) at La Silla Observatory (Chile), the  NASA Infrared Telescope Facility (NASA-IRTF), located at the Mauna Kea observatory in Hawaii (USA), and the Telescopio Nazionale Galileo (TNG), located at the Roque de los Muchachos Observatory in La Palma, Canary islands (Spain). The observational conditions are reported in Table~\ref{observations}.

\subsection{Visible}
The visible data presented in this work were obtained during three observational campaigns (June 2005, October 2005 and November 2006) at NTT, and in December 2012 at TNG.
At TNG spectra were acquired using the DOLORES detector (Device Optimised for the LOw RESolution), with the low resolution blue (LR-B) grism, spanning the 0.38-0.80 $ \mu m $ range,
and a 2 arcsec slit width, oriented along the parallactic angle to reduce the effects of atmospheric differential refraction.
At NTT, spectra were acquired using  EMMI (ESO Multi Mode Instrument), equipped with 
the grism \#1 and the 5 arcsec slit in RILD (red) mode, to cover the 0.41 - 0.96 
 $\mu m$ range.

Data reduction was performed with the software package ESO-Midas using standard procedures, as described in Dotto et al. (2006), Fornasier et al. (2004), and Perna et al. (2010). The basic procedure for converting raw flux measurements to reflectance spectra started with median bias subtraction and flat-field correction. Then, 2-D spectra were collapsed into mono-dimensional spectra, integrating the flux along the spatial axis, with an aperture radius generally between 1.5 and 2 FWHM and subtracting the sky contribution. We performed wavelength calibration identifying different spectral lines from  Ne-Ar and He-Ar calibration lamps. We also applied the extinction law for each site  and performed  extinction correction. The reflectivity was obtained by dividing the spectrum of each asteroid by the one of the solar analogue (usually the one with the closest airmass of the object). Spectra were smoothed with a median-filtering technique where, within a ten-pixel box, the original value was replaced by the median one if the two differed by a set threshold of 10\%. Finally spectra were normalised at 0.55 $\mu m$.

\subsection{Near-infrared}
The near-infrared spectra were obtained during three runs at IRTF (August, September, and October 2004), two at TNG (August 2006 and July 2007), and one at NTT in October 2006; we used IRTF equipped with the SpeX instrument (Rayner et al. 2003), a low-resolution prism, with a wavelength range from 0.8 to 2.5 $\mu m$ and a dispersion of 50 \textup{\AA} per pixel. We used an 0.8 arcsec slit, oriented along the parallactic angle, to reduce atmospheric effects.
At TNG we used the Near Infrared Camera Spectrometer (NICS), coupled with the AMICI prism, in low-resolution mode. We covered the 0.9 - 2.4 $\mu m$ spectral range and used a 2 arcsec slit, always oriented along the parallactic angle.
At NTT, spectra were collected using the SofI (Son OF Isaac) instrument in the low-resolution mode, using the blue (0.95  - 1.64 $\mu m$) and red grisms (1.53 - 2.52 $\mu m$), and then combining the overlapped region.

NIR observations were carried out using the standard technique of moving the objects along the slit between two positions denoted A and B. Cycles are repeated on ABBA sequences until the total exposure time is reached.

Standard procedures were used in the reduction process, as before with the visible data. We corrected spectra for flat field; then, bias and sky removal were obtained by subtracting the two spectra and producing $A - B$ and $B - A$ frames. We combined and averaged all pairs to obtain the final spectrum, which was then extracted. IRTF and NTT spectra were calibrated in wavelength through spectral lines from different calibration lamps (Ne, Ar).
For the TNG-NICS, spectral wavelength calibration was obtained using a look-up table available on the TNG website\footnote{ http://www.tng.iac.es/instruments/nics/spectroscopy.html}.
Extinction and solar removal was carried out by dividing the spectrum of each asteroid by the best solar analogue, generally the one with the closest airmass or the one observed right after or before the asteroid.
The NIR spectra of the observed asteroids were normalised at 1.25 $\mu m$. \\

\object{}

For asteroids \object{(13553) Masaakikoyama}, \object{(68359) 2001 OZ13}, \object{(203471) 2002 AU4}, and \object{(277127) 2005 GW119}, we combined our visible data with infrared spectra provided by the MIT-UH-IRTF Joint Campaign for NEO Reconnaissance\footnote{http://smass.mit.edu/minus.html}.
The spectra of the asteroids with infrared and visible range were overlapped, merging the common wavelength region using good atmospheric transmission zones to retrieve the normalisation factor, following procedures used by Fornasier et al. (2008).

{}

\section{Taxonomic classification and meteorite links}
We performed a $\chi^2$ best-fit method between
 our spectrum and the most representative one of each class of the Bus-DeMeo taxonomy (DeMeo et al. 2009), an extension in the infrared of the SMASSII spectral taxonomy (Bus \& Binzel, 2002), spanning a wavelength range between 0.45 to 2.45 $\mu m $ and finally counting 24 classes. The results of our taxonomic classification are reported in Table \ref{taxonomy}. Eleven asteroids are classified as belonging to the S-complex, while only two NEAs are classified as belonging to the C-complex. Nine of these objects were classified for the first time.

We investigated the suitable meteorite analogue type by comparing
the acquired reflectance spectra with laboratory reflectance spectra of meteorites.
Such a comparison presents some limitations (Gaffey et al. 2002) associated with the spectral variations due to changes 
in grain size, viewing geometry, etc. Nonetheless, all of these effects mainly produce a 
variation in the absorption band contrast, but do not significantly affect the wavelength positions of the absorption 
features, which are diagnostic for surface composition  and therefore give us constraints on it.

Using the M4AST online tool (Popescu et al. 2012), we automatically used a $\chi^2$  method
to compare our observed spectra with the whole sample of meteorite spectra included in the
RELAB database. 
 For each observed target, we analysed the 50 fits by visual inspection with the lowest $\chi^2$ value given by the tool: to identify the suitable meteorite analogue of each observed NEO amongst 
them,
we checked the presence and position of diagnostic spectral features, we took the meteorite and NEO
albedo into account when available, and we considered the overall spectral match with each potential meteorite analogue.
By comparing the spectral parameters of our observed bodies with meteorite spectral parameter ranges (see e.g. Clark et al. 2011 
and Dunn et al. 2010), we ended up with a few meteorites belonging to the same  group, which can be considered as suitable spectral analogues. 
For each target we suggest a possible meteorite analogue characterised by a low $\chi^2$ value and an overall good-quality fit.
Amongst the different meteorites giving similar $\chi^2$ values, we preferred 
the matches that mimicked diagnostic spectral features over matches that did not. 
Table \ref{taxonomy} reports, for each observed NEO, the meteorite that better reproduces the NEO spectral behaviour, 
together with its computed $\chi^2$.
The matches we suggest are also shown in Figs. \ref{c-type} - \ref{stype2}. 

\subsection{The C-complex}

   \begin{figure}
   \centering
   \includegraphics[angle=0,width=9cm]{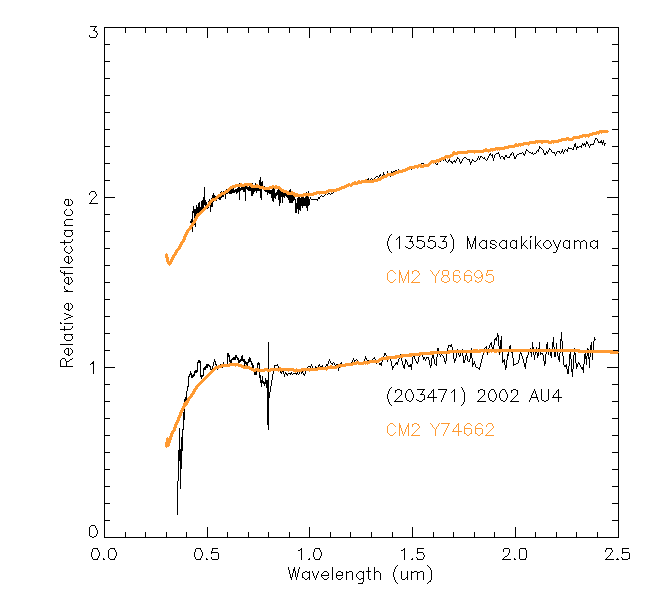}
      \caption{Visible spectra of C-complex asteroids \object{(13553) Masaakikoyama} and \object{(203471) 2002 AU4} with their meteorite analogues, combined with the NIR spectra made available by the MIT-UH-IRTF Joint campaign for NEO Reconnaissance.
The spectra are normalised at 0.55 $\mu$m and shifted for clarity.}
        \label{c-type}
   \end{figure}

In the Bus-DeMeo taxonomy, C-complex classes have weak spectral features. Cg and Cgh-classes for example are characterised by a strong UV drop-off, but the Cg does not have a 0.7 $\mu m$ feature, which also appears in the Ch (See Bus (1999) and DeMeo et al. (2009)). In our survey only two asteroids seem to belong to the C group: \object{(13553) Masaakikoyama} has a classification of a Cg-type, while \object{(203471) 2002 AU4} seems to belong to C/Cg-type (Fig. \ref{c-type}).
 
Krugly et al. (2006) found for {\bf \object{(13553) Masaakikoyama}} a rotation period of 38 hours. Reddy et al. (2006) observed this asteroid in the near infrared and found an excess of thermal emission and estimated an albedo of 0.03 $\pm$ 0.01, in agreement with a carbonaceous composition. They also suggest an association with CV carbonaceous types. 
The visible spectrum we acquired, combined with the NIR spectrum taken from the MIT-UH-IRTF Joint Campaign, is
flat without any features and with a possible UV drop-off below 0.5 $\mu m $, suggesting a Cg-type classification. 
On the basis of the analysis of spectral parameters and the overall spectral behaviour, we
found agreement with CK and CM carbonaceous chondrite meteorites;
also taking the low albedo of the asteroid into account, 
which is more compatible with a CM classification, we suggest the CM2 Y86695 as meteorite
analogue.

The visible spectrum we obtained for {\bf \object{(203471) 2002 AU4}} is flat and slightly bluish, with a possible 
feature around 0.75 $\mu m $.  
We classified 2002 AU4 as a C or a possible Cg-type. We also found a good 
match with the  spectra of CM carbonaceous chondrite meteorites. 
In Fig. \ref{c-type} we report the observed spectrum and the one of the CM2 meteorite Y74662.

\subsection{The S-complex: S, Sa, Sv, Sr-types}

  \begin{figure}
   \centering
   \includegraphics[angle=0,width=9cm]{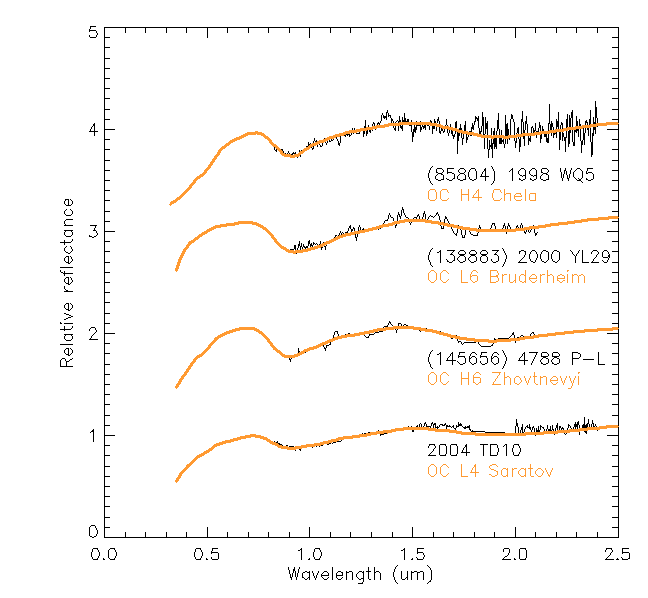}
      \caption{NIR spectra of S-complex asteroids \object{(85804) 1998 WQ5}, \object{(138883) 2000 YL29}, \object{(145656) 4788P-L}, and \object{2004 TD10} with their meteorite analogues. The spectra are normalised at 1.25 $\mu$m and shifted for clarity. }
         \label{stype3}
   \end{figure}

Objects that reside in the S-complex are identified by the 1 and 2 $\mu m $ features with shallow absorption bands. There is also a subtle classification (Sa, Sr, Sq, Sv) according to the exact depth and position of spectral bands.
When spectra are extended in the near-infrared, many S-objects became less and less separated.
In our survey two asteroids belong to the Sv class, \object{(85804) 1998 WQ5} and \object{(145656) 4788 P-L}; one object exhibits a Sr spectrum: \object{(138883) 2000 YL29}; one remains in the original S-class: \object{2004 TD10 }(Fig. \ref{stype3})

Oey (2006)  found a rotation period of P = (3.0089 $\pm$ 0.0001) h for asteroid {\bf \object{(85804) 1998 WQ5}}, and 
Mainzer et al. (2011) estimated an albedo of  0.24. 
Its spectrum shows a minimum around 0.9 $\mu m $ and a shallow band at 2 
$\mu m$. We classified it as an Sv-type and found general agreement  
with the spectral parameters of H ordinary chondrite meteorites. 
The spectrum of the ordinary chondrite H4 Chela is able to reproduce the
general spectral behaviour and to match the band around 0.9 $\mu m$.

Thomas et al. (2011) find an albedo of 0.19 for {\bf \object{(138883) 2000 YL29}} and 
suggest an Sq-type classification.  The NIR spectrum we observed is very red 
in the 0.9-1.4 $\mu m $ region, with the presence of a possible absorption 
band at 0.9 $\mu m $ and another around 2.0 $\mu m $.  We suggest a 
possible Sr-classification.  Our observed spectrum exhibits general 
agreement with L ordinary chondrite meteorites. 
In Fig. \ref{stype3} we show it with the one of the 
ordinary chondrite L6 Bruderheim, which gives the lowest $\chi^2$ value and is 
able to reproduce the general spectral 
behaviour.

The NIR spectrum we acquired for {\bf \object{(145656) 4788 P-L}} is somewhat red, and it shows a possible absorption band at 1.0 $\mu m $ and another around 1.9 $\mu m $. 
We classified it as an Sv-type.  On the basis of comparison of our 
spectrum with laboratory spectra of meteorites in the 
RELAB database, we suggest a possible link with 
H ordinary chondrite meteorites;  in particular, we show the comparison with the H6 Zhovtnevyi in Fig. \ref{stype3}.  Their spectral parameters match the general spectral behaviour of this NEA.

We observed {\bf \object{2004 TD10}} two weeks after the discovery.  
The spectrum has a high S/N, but the regions around 1.4 $\mu m$ 
and around 1.9 $\mu m$ were removed due to the high residuals of atmospheric bands.
The S-type nature of this asteroid is suggested by the two absorption bands: 
a clear one at 1 $\mu m $, and a shallower one at 2 $\mu m $.
We found general spectral agreement with L ordinary 
chondrite meteorites like the L4 Saratov shown in Fig. \ref{stype3}, which has spectral behaviour that is very similar to this NEA.

\subsection{The S-complex: Sq/Q types}

\begin{figure}
   \centering
   \includegraphics[angle=0,width=9cm]{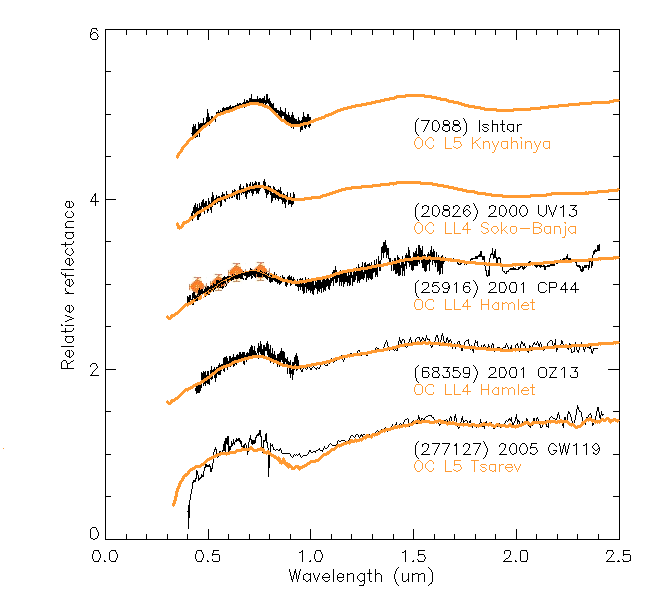}
      \caption{Visible-NIR spectra of S-complex asteroids \object{(7088) Ishtar}, \object{(20826) 2000 UV13}, \object{(25916) 2001 CP44}, \object{(68359) 2010OZ13}, and \object{(277127) 2005GW119} with their meteorite analogues. For \object{(68359) 2010OZ13 }and \object{(277127) 2005GW119}, we combined our visible spectra with the NIR data available in the MIT-UH-IRTF public database. The spectra are normalised at 0.55 $\mu$m and shifted for clarity.}
         \label{stype1}
   \end{figure}

\begin{figure}
   \centering
   \includegraphics[angle=0,width=9cm]{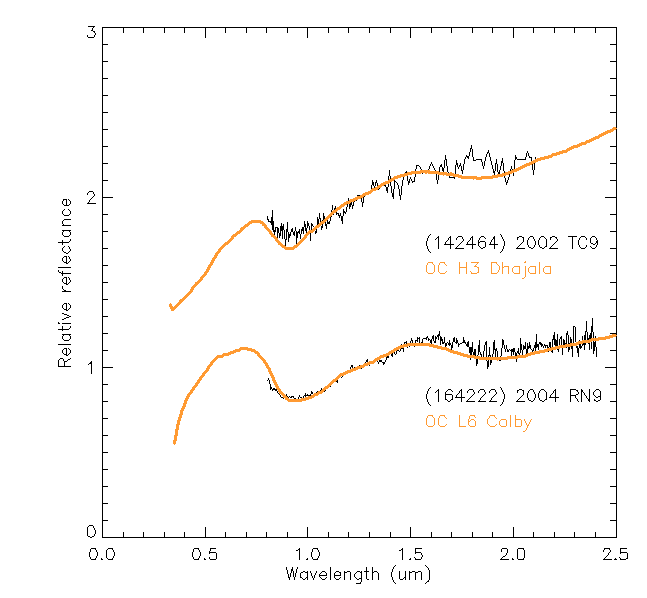}
      \caption{NIR spectra of S-complex asteroids \object{(142464) 2002 TC9} and \object{(164222) 2004 RN9} with their meteorite analogues. The spectra are normalized at 1.25 $\mu$m and shifted for clarity.}
         \label{stype2}
   \end{figure}

In the near-infrared, Q and Sq types are very similar to each other, with a broadened and rounded 1 $\mu m $ band that reaches a minimum reflectance level around   0.9-0.95 $\mu m $ and a band around 2 $\mu m $, similar to olivine and pyroxene assemblages. Q-types show a moderately steep UV slope shortwards of 0.7 $\mu m $, Sq-types can contain a relatively strong feature around 0.63 $\mu m $, and some of them exhibit also another feature around 1.3 $\mu m $. In our survey seven objects are classified as Sq types: \object{(25916) 2001 CP44}, \object{(68359) 2001 OZ13},  \object{(164222) 2004 RN9}, \object{(277127) 2005 GW119}, \object{(7088) Ishtar}, \object{(20826) 2000 UV13}, and \object{(142464) 2002 TC9} (Figs. \ref{stype1} and \ref{stype2}). For the last three, a classification as Sr type cannot be excluded.

The object {\bf \object{(7088) Ishtar}} is a binary asteroid  with a slope parameter G = 0.24 (Pravec et al. 2012).
Reddy et al. (2007) used lightcurve analysis to obtain an estimate of the synodic orbital period of (20.63$\pm$0.02) h. They also made the assumption that the primary has an a spheroidal shape, giving its rotation period of P = (2.6786$\pm$0.0002) h and a lightcurve amplitude of 0.11 mag.
The visible spectrum we obtained is very red until 0.75 $\mu m $  
and shows a prominent band around 0.95 $\mu m $, suggesting an Sq or Sr-type 
classification. However, owing to the binary nature of this asteroid, 
rotationally resolved spectra obtained in different geometrical view 
are needed to distinguish between possibly different compositions of the pair.
 We found general agreement between our observed spectrum and those of 
ordinary chondrite meteorites belonging 
to the L class. In Fig. \ref{stype1} we show the comparison with  
laboratory spectrum of  
the L5 ordinary chondrite Knyahinya, which gives the lowest $\chi^2$ value and matches the overall spectral behaviour.

The asteroid {\bf \object{(20826) 2000 UV13}} 
was observed by Binzel et al. (2004), who classified it as belonging to the Sq class. It has an estimated albedo of 0.18 (Thomas et al. 2011) and a rotation period of 12 hours\footnote{ http://www.asu.cas.cz/~ppravec/newres.txt}. 
The visible spectrum we obtained is red until 0.75 $\mu m $ 
and shows a possible 0.9 $\mu m $ band, but infrared data are needed to confirm it. It could be either classified as an Sq or Sr-type. The reflectance spectrum  agrees with those of LL meteorites: we found a match with the LL4 Soko-Banja, whose  
spectrum is shown in Fig. \ref{stype1}.

\textbf{ } 
Based on images taken during three different nights,
Elenin \& Molotov (2012) find a rotation period for {\bf \object{(25916) 2001 CP44}} 
of P= (4.19$\pm$ 0.01) h and a light curve amplitude of 0.28 mag. Thomas et al. (2011) classified this object as a Q-type with an albedo of 0.21. In the visible range we observed this object with EMMI, using both the imaging and spectroscopic modes. Our photometric data allowed us to obtain the colours B - V = (0.819$\pm$0.020), V - R = (0.463$\pm$0.013), and V - I = (0.789$\pm$0.013). The spectrum shows a moderately red  slope.  
The NIR spectrum we obtained has two absorption features at 1 and 2 $\mu m $, which are typical of the silicate asteroids, in particular of an Sq-type.
Meteorites that reproduce the observed spectral behaviour better are those
belonging to the LL group of ordinary chondrites. In particular, we report the spectrum of  LL4 ordinary chondrite Hamlet in 
Fig. \ref{stype1}.

Within the ExploreNEO survey, Thomas et al. (2011) estimated 
an unusual albedo of $0.47^{+0.32}_ {-0.22}$  and an Sq spectral type for {\bf \object{(68359) 2001 OZ13}}. 
We acquired two visible spectra forty minutes away from each other. No sign of rotational variation has been found:
both spectra appear very red until 0.78 $\mu m $,  
and afterwards there is the 0.9 $\mu m $ absorption band. Our spectrum, combined with the NIR data available in the MIT-UH-IRTF Joint Campaign, confirm its classification as an Sq-type. The best meteorite  analogues we found belong to the LL class. In particular, we show the match we obtained with the ordinary 
chondrite LL4 Hamlet in Fig. \ref{stype1}.

Combining the visible spectrum we observed for {\bf \object{(277127) 2005 GW119}} with the 
infrared spectrum available in the ,literature (MIT-UH-IRTF Joint Campaign) we 
classified it as an Sq-type. 
 Ordinary chondrites belonging to the L class match the observed 
spectrum, but there is some discrepancy in the 1 $\mu m $ region, where the 
meteorite spectra generally lead to underestimating the observed one. In
Fig. \ref{stype1} we report the spectrum, along with the one of the ordinary chondrite L5 Tsarev, which is able to better reproduce the diagnostic spectral feature at 1 $\mu m$.

Mainzer et al. (2011) find an albedo of 0.12 for {\bf \object{(142464) 2002 TC9}}. 
The NIR spectrum we obtained suggests this asteroid is an Sq or Sr-type, with an absorption 
band at 1 $\mu m $ but a reddish slope behaviour around 2 
$\mu m $. 
We found general agreement with H ordinary chondrite meteorites, and 
in Fig. \ref{stype2} we report our comparison with the H3 Dhajala,  whose spectrum matches the
wavelength position of the diagnostic 1 $\mu m$ band better.

The object {\bf \object{(164222) 2004 RN9}} has a NIR spectrum that shows absorption bands around 
0.9 $\mu m $ and 1.9 $\mu m $, in agreement with an Sq-type classification. 
 L-ordinary chondrites generally agree with the overall spectral behaviour of the asteroid; in particular, we chose the spectrum of the L6 Colby, which matches the wavelength position of the 1 and 2 $\mu m$ band (Fig. \ref{stype2}) better.

\section{Mineralogy analysis}

As mentioned before, S-type asteroids exhibit a generally red spectrum with prominent bands in the NIR region, typically of pyroxene and olivine assemblages, around 1 and 2 $\mu m $. These absorption bands are caused by the presence of $Fe^{2+}$ cations located in the M2 crystallographic site (Adams, 1974, 1975; Burns, 1993).

There are multiple parameters that are usually compared in the literature to infer the mineralogy properties of silicate asteroids, such as band minimum, band centres (BI and BII), band depth, band width, and band area ratio (BAR, between the 2 and the 1 $\mu m $ bands) (Gaffey et al. 2002). The ones that are used the most are band minimum and centre, the lowest point of the reflectance spectrum, before and after continuum removal, respectively. The continuum in this case is defined as the linear slope between the relative maxima, generally located around 0.7 and 1.4 $\mu m $.

\textbf{}
To compute band minimum we fitted a six-degree polynomial over the bottom third of each band. Errors were computed using a Monte Carlo simulation, randomly sampling data 100 times  and taking the standard deviation as incertitude. Theoretically, to determine the band centre, we should subtract the continuum properly from the asteroid's spectrum.
But to apply this technique, we should have the visible and NIR spectra taken at the same time, since different observation conditions could result in markedly different spectral slopes, hence the subsequent band centre determination. 
For this reason, we preferred to compute band minima and use laboratory determinations to derive band centres. We used the simple relation between band minimum and band centre described by Cloutis \& Gaffey (1991)

\begin{equation}
BI = BI_{min} + 0.007 \, \mu m \, \pm 0.004
.\end{equation}

Since the start position for Band I and the end position for Band II are not always available in our spectra, to compute the band areas we fitted a six-degree polynomial over each band by considering the starting point around 0.7 $\mu m $ for Band I and for Band II the furthest point around 2.5 $\mu m $.
We computed the band area first by dividing out the continuum and then measuring the area between the new continuum and the polynomial fit for the 1 and 2 $\mu m $ bands. BAR was subsequently obtained by dividing Band II area by  Band I area.

BAR measurements are also affected by temperature shifts, so to compare our measurements with laboratory data, we used an inverse correlation found by Sanchez et al. (2012) for different assemblages of olivine and orthopyroxene, e.g. BAR correction decreases when the temperature increases. 
We established an approximate surface temperature using a standard thermal model (Lebofsky \& Spencer, 1989):

\begin{equation}
 T = [(1-A)L_{\sun} /16 \, \epsilon \eta  \sigma \pi  r^2]^{1/4}.
\end{equation}

\noindent 
where A stands for the bond albedo of the asteroid, $ L_{\sun}$ is the solar luminosity, 
$\sigma$  the Boltzman constant, and r  the distance from the Sun in metres. For our computations we assumed that the beaming factor $ \eta$ is approximately 1 and the infrared emissivity $\epsilon \sim$ 0.9. The final temperature however is not particularly sensitive to minor changes in these parameters or to changes in the albedo. For objects with unknown albedo we considered the mean one of their taxonomical classes  (0.15 for an S-type, 0.04 for a C-type), as taken from Shestopalov \& Golubeva (2011).
Once we found an approximate temperature for the surface of the asteroids, we applied Sanchez's relation to retrieve BAR correction:

\begin{equation}
BAR_{corr} = 0.00075 \times T (K) - 0.23.
\end{equation}

We inferred mineralogy properties by applying formulas derived by Dunn et al. (2010) and used also by Dunn et al. (2013) to study  the relationships between fayalite (Fa) in mafic minerals and ferrosilite (Fs) in pyroxene and spectral features. These equations were obtained for ordinary chondrites and theoretically could be applied to asteroids that share a similar mineralogy. With these relationships it is possible to determine the molar content of fayalite in olivine, ferrosilite in pyroxene, and also the relative percentage of olivine and pyroxene.
The correlation between BI and the molar content of Fa and Fs can be expressed as a second-order polynomial fit:

\begin{equation}
Fa = -1284.9 \times (BI)^2 + 2656.5 \times (BI) -1342.3,\end{equation}

\begin{equation}
Fs = -879.1 \times (BI)^2 + 1824.9 \times (BI) -921.7,\end{equation}

\noindent
while the mineral  ratio between olivine and pyroxene can be calculated from BAR values, using the relationship 

\begin{equation}
ol/(ol+px) = -0.242 \times BAR + 0.728.
\end{equation}

\begin{figure}
   \centering
   \includegraphics[angle=0,width=10cm]{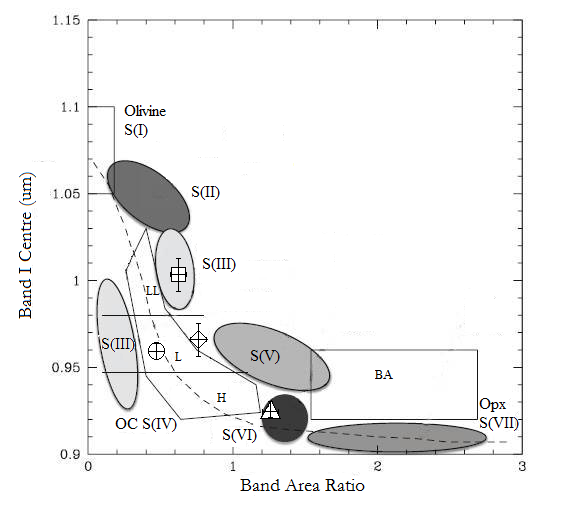}
      \caption{Plot of BI versus BAR for \object{(25916) 2001 CP44} (square), \object{(85804) 1998 WQ5} (triangle), \object{(164222) 2004 RN9} (circle), and \object{2004 TD10} (diamond). The rectangular zone includes pyroxene-dominated basaltic achondrite (BA) assemblages. The S(IV) subgroup represents mafic silicate components of ordinary chondrites (OC), from Gaffey et al. (1993). The dashed curve is the limit of the mixing line for olivine and orthopyroxene mixing, as shown in Cloutis \& Gaffey (1986). Horizontal lines separate H, L, and LL chondrites. Adapted from Gaffey et al. (1993).}
         \label{barband}
   \end{figure}

\begin{figure}
   \centering
   \includegraphics[angle=0,width=10cm]{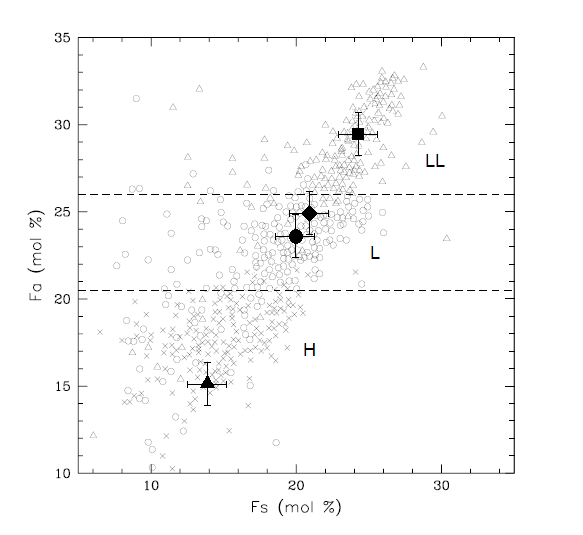}
      \caption{Molar content of Fa vs. Fs for NEAs \object{(25916) 2001 CP44} (square), \object{(85804) 1998 WQ5} (triangle), \object{(164222) 2004 RN9} (circle), and \object{2004 TD10} (diamond), along with values for LL (open triangles), L (open circles), and H (x) ordinary chondrites, adapted from Sanchez et al. (2013). The error bars correspond to the values determined by Dunn et al. (2010) 1.4 mol\% for Fs and 1.3 mol\% for Fa. The horizontal dashed lines represent the approximate boundaries for H, L, and LL chondrites.}
         \label{fafs}
   \end{figure}

\begin{figure}
   \centering
   \includegraphics[angle=0,width=9cm]{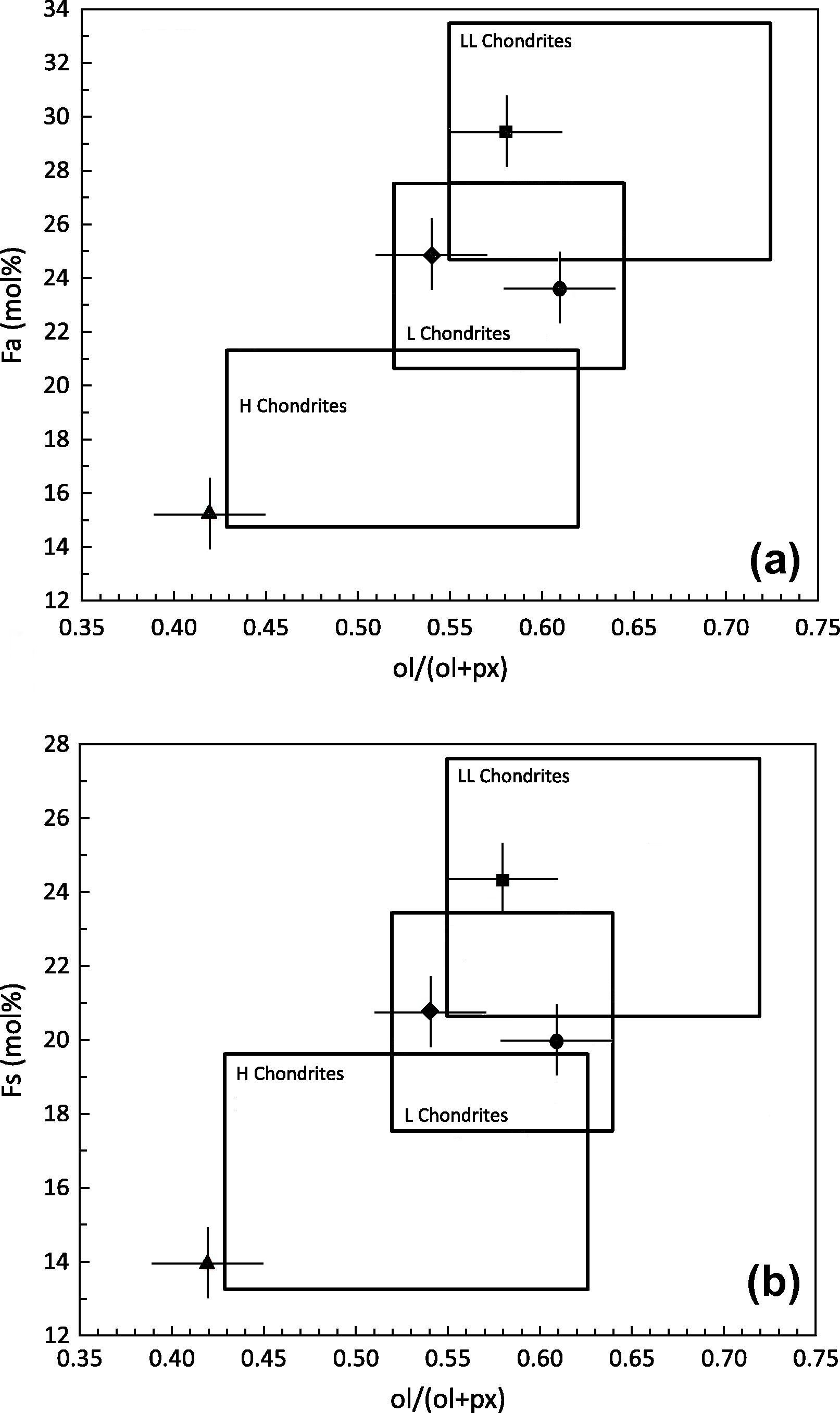}
      \caption{(a) Spectrally derived mol\% Fa and (b) mol\% Fs plotted as a function of derived ol/(ol+px) for \object{(25916) 2001 CP44 }(square), \object{(85804) 1998 WQ5} (triangle), \object{(164222) 2004 RN9} (circle), and \object{2004 TD10} (diamond). Solid boxes represent the range of spectrally derived ol/(ol+px) for H, L, and LL ordinary chondrites, together with error bars from Dunn et al. (2010). Error bars represent the least mean square of the errors (0.03 for ol/(ol+px), 1.3 mol\% for Fa, and 1.4 mol\% for Fs). Adapted from Dunn et al. (2013).}
         \label{olivine}
   \end{figure}

We were able to calculate BI and the BAR values for four objects (\object{2004 TD10}, \object{(25916) 2001 CP44}, \object{(85804) 1998 WQ5}, and \object{(164222) 2004 RN9)}. For  \object{(68359) 2001 OZ13}, \object{(142464) 2002 TC9}, and \object{(277127) 2005 GW119} we computed BI, though it was not possible to determine the BAR owing to the low S/N in the Band II region. 
The derived olivine and pyroxene chemistries  are shown in Figs.  \ref{barband}-\ref{olivine}. Figure \ref{barband} shows BAR vs BI for our targets, along with the Gaffey classification for olivine-orthopyroxene assemblages, where the S(IV) subgroup represents the mafic silicate components of ordinary chondrites. In Fig. \ref{fafs} we plot the molar content of Fs vs Fa, as found with the Dunn et al. (2010) relations. In Fig. \ref{olivine} the same values are plotted against the olivine and pyroxene ratio, along with compositional regions for H, L, and LL found by Dunn et al. (2010). Band centres, BARs and slopes are shown in Table \ref{magnitudes}.

{}

 In Fig. \ref{barband} \object{(25916) 2001 CP44} is
just above the polygonal region defined by Gaffey et al. (1993) for the ordinary chondrites, near the LL-chondrite region, and just above the mixing line of olivine-ortopyroxene, as described in Cloutis \& Gaffey (1986). 
Its values of Fa and Fs are consistent with those derived by LL ordinary chondrites by Dunn et al. (2010), which give a range for fayalite between $(26 - 32) \pm  1.3$ mol\%  and between $(22 - 26) \pm  1.4$ mol\% for ferrosilite. The relative olivine percentage estimated from BAR for this asteroid also agree with spectrally-derived measurements reported by Dunn et al. (2010) for LL ordinary chondrites (Fig. \ref{olivine}), which give ol/(ol+px) values in the  $(0.58-0.69)\pm  0.03$ range, and a specific value for the Hamlet chondrite of 0.59. We conclude that this asteroid could be consistent with an LL ordinary chondrite, in agreement with the best meteoritic analogue we found by spectral comparison.

The asteroid \object{(85804) 1998 WQ5} 
in Fig. \ref{barband} and in Fig. \ref{olivine} falls just outside the H-chondrite region.
Nevertheless, its pyroxene chemistry is located in the range estimated for H chondrites (Fig. \ref{fafs}), in agreement with the meteorite analogue we found, the ordinary chondrite H4 Chela. Despite some inconsistencies among Figs. \ref{barband}, \ref{fafs}, and \ref{olivine}, this asteroid could be coherent with an H chondrite.

The BI and the BAR of \object{(164222) 2004 RN9}  place it in the region corresponding to L ordinary chondrites (Fig.  \ref{barband}). This is in strong agreement with the meteoritic analogue we found, the L6 ordinary chondrite Colby.
Using the equations for molar contents,  we determined olivine and pyroxene chemistries  consistent with those derived by Dunn et al. (2010) for L ordinary chondrites (i.e. Fa between $(21 - 27) \pm  1.3$ mol\% and Fs between $(17 - 23)\pm  1.4$ mol\%). From the BAR value we estimated the composition of olivine and pyroxene, in agreement with spectrally-derived measurements reported by Dunn et al. (2010) for L ordinary chondrites, having a relative percentage of olivine between $(0.54 - 0.62) \pm  0.03$ (Fig. \ref{olivine}).

In 
Fig. \ref{barband} \object{ 2004 TD10} is just outside the bounds of the S(IV) region, in the L chondrites zone, in agreement with the meteoritic analogue that we found, the chondrite L4 Saratov. According to our calculations, the olivine and pyroxene chemistries are 
consistent for values for L chondrites derived by Dunn et al. (2010). The ol/(ol+px) ratio found for this asteroid also agrees with the Dunn et al. (2010) measurements for L ordinary chondrites.

\section{Conclusions}

We collected seven spectra in the visible and seven in the near infrared range from different telescopes (NASA-IRTF, ESO-NTT, TNG) for 13 NEAs with a low $\Delta V$.
These targets are all suitable candidates for future space missions. 

Through the analysis of the spectra obtained, we estimated surface composition of these asteroids, yielding classifications according to Bus \& DeMeo taxonomy. We found 11 S-complex objects and two C-complex objects. In particular, four objects are Sq-types, three  Sq/Sr-types, two Sv-types, and two exhibit a S or Sr spectrum, while there are two C/Cg-types. Through the comparison of the observed NEAs and the RELAB meteorite database, we obtained good matches between ordinary chondrites and S-type asteroids and between carbonaceous chondrites and C-type asteroids. Finally, we investigated mineralogy by sampling the prominent bands in the NIR, which is typical of olivine and pyroxene assemblages. We obtained good agreement with the previous meteorite analogues we found and with the overall results from Dunn et al. (2010).

Visible and NIR spectroscopic observations are fundamental for defining the physical characteristics of the NEO population and resolving the conundrum about the prevailing abundance among NEOs. The high rate of silicate asteroids, which until only recently were seen to be the dominant component of the NEO population, could be in principle a selection effect, as suggested by the NEOWISE survey (Masiero et al. 2011), or is more probably due to the high efficiency of the transport mechanisms in the inner main belt, where Q and S-type asteroids are most commonly found (Fieber-Beyer et al. 2012).

 The physical characterisation of an asteroid is generally the first step towards developing and designing any space mission to a target asteroid, such as a mitigation mission or a rendez-vous mission. Spectral characterisation of NEOs is a powerful tool for defining successful mitigation strategies in the case of possible impactors, since there is a high degree of diversity among the composition of NEO population. For a kinetic impactor mitigation mission, the stress waves produced by an impact propagate and attenuate differently in a C-type or in an S-type, since the whole process strongly depends on the asteroid's porosity.

Missions to NEOs are becoming more and more feasible, and ground-based spectroscopic surveys, like NEO-SURFACE, are a fundamental step in the broader framework of the NASA's FY2014 Asteroid Strategy, which as stated by President Obama in 2010, has the aim of sending a human mission to an asteroid by 2025. The selection of a future target would be strongly affected by the low $\Delta V$ and the peculiarities of its spectrum. We therefore need to observe and classify the largest possible number of these low $\Delta V$-asteroids.

\begin{table*}[p]
\caption{Observational circumstances.}
\label{observations}      
\centering
\begin{tabular}{l c l c c c c c c r}        
\hline\hline                 
Object                 & $\Delta V$ & Date & $\Delta$ &   r   & Telescope/ & Spectral & $T_{exp}$ &  Airmass & Solar Analogue \\
                       &   (km/s)   &      &   (AU)   & (AU)  & Instrument & range    &   (s)     &          & (airmass)     \\  
\hline                        
(7088) Ishtar          &  6.684 & 31/10/05 & 0.518    & 1.289 & NTT - EMMI    & V     & 400       &  1.6     & SA 93-101 (1.7) \\
(13353) Masaakikoyama  &  6.678 & 19/06/05 & 0.297    & 1.274 & NTT - EMMI    & V     & 300       &  1.2     & SA 107-684 (1.1) \\
(20826) 2000 UV13      & 10.452 & 20/06/05 & 1.942    & 2.411 & NTT - EMMI    & V     & 600          &  1.0     & HD 218251 (1.1)  \\
(25916) 2001 CP44      &  8.219 & 26/10/06 & 1.296    & 2.227 & NTT - SofI    & NIR   & 16 x 120  &  1.0     & SA 115-271 (1.0) \\
                       &        & 15/11/06 & 1.400    & 2.344 & NTT - EMMI    & V     & 360          &  1.1     & HD 2141 (1.2)  \\
(68359) 2001 OZ13      &  6.539 & 31/10/05 & 0.582    & 1.512 & NTT - EMMI    & V     & 300          &  1.1     & HD 2141 (1.1)  \\
                       &        & 31/10/05 &          &       & NTT - EMMI    & V     & 600          &  1.2     & SA 98-978 (1.1)  \\
(85804) 1998 WQ5       &  9.541 & 23/08/04 & 1.094    & 1.997 & IRTF - SpeX   & NIR   & 14 x 120  &  1.1     & SA 112-1333 (1.1) \\
(138883) 2000 YL29     &  7.633 & 15/07/07 & 0.700    & 1.069 & TNG - NICS    & NIR   & 28 x 120  &  1.4     & HD 148642 (1.8)  \\
(142464) 2002 TC9      &  6.934 & 19/08/06 & 0.352    & 1.269 & TNG - NICS    & NIR   & 12 x 120  &  1.0     & SA 115-271 (1.0) \\
(145656) 4788 P-L      &  7.401 & 15/07/07 & 0.312    & 1.156 & TNG - NICS    & NIR   &  8 x 120  &  1.2     & HD 144873 (1.1)  \\
(164222) 2004 RN9      &  6.436 & 23/09/04 & 0.113    & 1.114 & IRTF - SpeX   & NIR   & 16 x 120  &  1.1     & HD 1368 (1.1)  \\
(203471) 2002 AU4      & 10.025 & 20/12/12 & 0.103    & 1.065 & TNG - DOLORES & V     & 780       &  1.2     & SA 115-271 (1.1)  \\
(277127) 2005 GW119    &  6.063 & 20/12/12 & 0.325    & 1.290 & TNG - DOLORES & V     & 1800      &  1.2     & SA 115-271 (1.1)  \\
2004 TD10              & 10.127 & 21/10/04 & 0.049    & 1.028 & IRTF - SpeX   & NIR   & 50 x 120  &  1.6     & HD 218251 (1.6)  \\
\hline                                   
\end{tabular}
~\\
\raggedright
\smallskip
NOTE: $\Delta$ and $r$  are the topocentric and the heliocentric distance, respectively.\\
\end{table*}

\begin{table*}[p]
\caption{Physical characteristics of the targets: absolute magnitude, albedo, diameter, taxonomical class, suitable meteorite analogue 
class, best match analogue, and $\chi^2$ value. }
\label{taxonomy}      
{\centering
\begin{tabular}{l c c c c c l c }        
\hline\hline                 
Object                     & H & $\rho_V$ & D$^1$ & Tax.     & Meteorite  & Meteorite           & $\chi^2$  \\
                           &   &          & (km)  & class    & class & name           &           \\
\hline
{\bf(7088) Ishtar}         &17.08& ...  & ...  & {\bf Sq/Sr} & OC L  & L5 Knyahinya   &   $3.20 \times 10^{-4}$\\
{\bf(13353) Masaakikoyama} &16.71& 0.03 & 3.49 & {\bf Cg}    & CC CM & CM2 Y86695     & $1.14 \times 10^{-3}$\\
(20826) 2000 UV13          &13.70& 0.18 & 5.70 & Sq/Sr       & OC LL & LL4 Soko-Banja &$2.60 \times 10^{-4}$ \\
(25916) 2001 CP44          &13.40& 0.21 & 6.06 & Sq          & OC LL & LL4 Hamlet     & $1.14 \times 10^{-3}$\\
(68359) 2001 OZ13          &17.80& 0.47 & 0.53 & Sq          & OC LL & LL4 Hamlet     & $ 7.20 \times 10^{-4}$\\
{\bf(85804) 1998 WQ5}      &15.30& 0.24 & 2.36 & {\bf Sv}    & OC H  & H4 Chela       &$ 5.90 \times 10^{-4}$ \\
(138883) 2000 YL29         &16.70& 0.19 & 1.39 & Sr          & OC L  & L6 Bruderheim  & $7.80 \times 10^{-4}$\\ 
{\bf(142464) 2002 TC9}    &18.10& 0.12 & 0.92 & {\bf Sq/Sr} & OC H  & H3 Dhajala     &3.03 $ \times 10^{-3}$\\
{\bf(145656) 4788 P-L}    &16.35& ...  & ...  & {\bf Sv}    & OC H  & H6 Zhovtnevyi  & $1.19 \times 10^{-3}$\\
{\bf(164222) 2004 RN9}    &20.66& ...  & ...  & {\bf Sq}    & OC L  & L6 Colby       & $6.20 \times 10^{-4}$\\ 
{\bf(203471) 2002 AU4}     &19.20& ...  & ...  & {\bf C/Cg}  & CC CM & CM2 Y74662     &3.14 $ \times 10^{-3}$\\  
{\bf(277127) 2005 GW119}   &18.49& ...  & ...  & {\bf Sq}    & OC L  & L5 Tsarev      & $8.10 \times 10^{-3}$\\   
{\bf2004 TD10}             &22.03& ...  & ...  & {\bf S}     & OC L  & L4 Saratov     & $3.10 \times 10^{-4}$ \\
\hline                        

\hline                                   
\end{tabular}
~\\
\smallskip
}  
NOTE: References. (1) Fowler \& Chillemi, 1992. \\
 In bold asteroids taxonomically classified for the first time.\\
\end{table*}

\begin{table*}[p]
\caption{Visibile slopes, band centres, BARs, and mineralogical composition.}
\label{magnitudes}      
\centering
\begin{tabular}{l l c c c c c c}        
\hline\hline                 
Object                           & Slope$^a$ (\%$/10^3$ A)       & BI ($\mu m$)             & BAR              & BAR corr              & Fa     (mol\%)         & Fs   (mol\%)     & Ol / ol + opx       \\
\hline                        
(7088) Ishtar                    & $12.24  \pm 0.03$ &     $0.952  \pm 0.003 $      & ... & ...            & ...            & ...            &...  \\
(13353) Masaakikoyama     & $4.27  \pm 0.17$ &     ...      & ... & ...            & ...            & ...            &...  \\
(20826) 2000 UV13             & $10.37  \pm 0.31$ &    ...      & ... & ...            & ...            & ...            &...  \\
(25916) 2001 CP44            & $10.00  \pm 0.23$ &    $1.003  \pm 0.009 $      & $0.72  \pm  0.05$ &  $0.62  \pm 0.05$           &  $29.5          \pm 1.3 $&$  24.3  \pm   1.4     $ &$0.58 \pm 0.03$\\
(68359) 2001 OZ13            & $16.13  \pm 0.35$ &    $0.992  \pm 0.005 $       & ... & ...            & ...            & ...            &...  \\
(85804) 1998 WQ5             & ...                      &    $0.924  \pm 0.005 $     & $1.35  \pm 0.05 $ &  $1.27  \pm 0.05 $           &  $15.3  \pm    1.3    $ & $13.9 \pm  1.4     $ &$0.42 \pm 0.03  $\\
(138883) 2000 YL29              & ... &    ...     & ... &...            & ...            & ...            &...  \\
 (142464) 2002 TC9             & ... &   $0.925  \pm 0.004$     & ... & ...            & ...            & ...            &...  \\
 (145656) 4788 P-L            & ... &    ... & ... & ...            & ...            & ...            &...  \\
 (164222) 2004 RN9             & ... &    $0.959  \pm 0.005 $     & $0.52  \pm 0.05$ &  $0.48  \pm 0.05$           & $ 23.6  \pm     1.3   $ & $19.9         \pm 1.4 $&$0.61 \pm 0.03 $\\
(203471) 2002 AU4              & $0.07  \pm 0.16$ &    ...      & ... & ...            & ...            & ...            &...  \\
(277127) 2005 GW119           & $13.23 \pm 0.97$ &   $0.937  \pm 0.003 $    & ... & ...            & ...            & ...            &...  \\
2004 TD10                        & ... &    $0.966  \pm 0.010$     & $0.79  \pm 0.05$ &  $0.76  \pm 0.05$           &  $24.9   \pm  1.3     $ & $20.8\pm 1.4 $&$0.54 \pm 0.03 $\\
\hline                                   
\end{tabular}
~\\
\raggedright
\smallskip
\small
NOTE: $^a$ Computed between 0.5-0.75 $\mu m$.
\end{table*}

\begin{acknowledgements}
We would like to thank all the telescope operators for their contribution. In this article we make use of meteorite spectra taken
with the NASA RELAB facility at Brown University. Part of the data utilised in this publication were obtained and made available by the  MIT-UH-IRTF Joint Campaign for NEO Reconnaissance. The IRTF is operated by the University of Hawaii under Cooperative Agreement no. NCC 5-538 with the National Aeronautics and Space Administration, Office of Space Science, Planetary Astronomy Program. The MIT component of this work is supported by NASA grant 09-NEOO009-0001 and by the National Science Foundation under Grants Nos. 0506716 and 0907766. Part of this work was supported by INAF (PRIN-INAF 2009 “Near Earth  
Objects”) and ASI (A.A. ASI I/079/9/0) funds. We would like also to thank the anonymous referee, for the thoughtful suggestions and comments, which helped us improve the manuscript.
\end{acknowledgements}


\begin{thebibliography}{}

   \bibitem[]{Adams74} Adams, J. B. 1974, JGR, 79, 4829
   \bibitem[]{Adams75}  Adams, J. B. 1975,  in Infrared and Raman spectroscopy of lunar and terrestrial minerals. Academic Press, San Diego, 91.
     \bibitem[]{Binzel04} Binzel, R. P., Rivkin, A. S., Stuart, J. S. et al. 2004, Icarus, 170, 259
   \bibitem[]{Burns} Burns, R. G. 1993, in  Mineralogical applications of crystal field theory, Cambridge univ. press, Cambridge
   \bibitem[]{Bus99} Bus, S.J. 1999, PhD Thesis, MIT
   \bibitem[]{Bus02}  Bus, S.J. \& Binzel R. P. 2002, Icarus, 158, 146
 \bibitem[]{Clark11}  Clark, B. E., Binzel, R. P., Howell, E. S. et al. 2011, Icarus, 216, 462
   \bibitem[]{cloutis86} Cloutis, E. A. \& Gaffey, M. J., Jackowski, T. L., Reed, K. L. 1986, JGR, 91, 11641
   \bibitem[]{cloutis91} Cloutis, E. A. \& Gaffey, M. J. 1991, JGR, 96, 22809
   \bibitem[]{} DeMeo, F. E., Binzel, R. P., Slivan, S. M., Bus, S. J. 2009, Icarus, 202, 160
 \bibitem[]{} Dotto, E., Fornasier, S.,  Barucci, M. A. et al. 2006, Icarus 183, 420
   \bibitem[]{} Dunn, T. L., McCoy, T. J., Sunshine, J. M., McSween, H. Y. 2010,  Icarus 208, 789
  \bibitem[]{} Dunn, T. L., Burbine, T. H., Bottke, W. F., Clark, J. P. 2013,  Icarus 222, 273
   \bibitem[]{} Elenin, L. \& Molotov I. 2012, MPBu, 39, 101
  \bibitem[]{} Fieber-Beyer, S. K., Gaffey, M. J., Hardersen, P. S., Reddy, V. 2012, Icarus, 221, 593
 \bibitem[]{} Fornasier, S., Dotto, E., Barucci, M. A., Barbieri C. 2004, A\&A, 421, 353
  \bibitem[]{} Fornasier, S.,  Migliorini A., Dotto E., Barucci, M. A. 2008, Icarus, 196, 119
\bibitem[]{} Fowler, J. W. \& CHillemi, J.R. 1992, in The IRAS Minor Planetary Survey, Philips Laboratory Hanscom Air Force Base, Massachussets.
   \bibitem[]{} Gaffey, M. J., Burbine, T. H., Piatek, J. L. et al. 1993, Icarus, 106,573 
   \bibitem[]{} Gaffey, M. J., Cloutis, E. A., Kelley, M. S., Reed, K. L 2002 in Asteroids III, ed. W. F. Bottke Jr., A. Cellino, P. Paolicchi, and R. P. Binzel, University of Arizona Press, Tucson, 183
\bibitem[]{}  Krugly, Yu. N., Gaftonyuk, N. M., Belskaya, I. N. et al. 2007, IAUS, 236, 385
\bibitem[]{} Lebofsky, L. A., \& Spencer, J. R. 1989, in Asteroids II, University of Arizona Press, Tucson, 128
\bibitem[]{} Mainzer, A., Grav, T., Bauer, J. et al. 2011, ApJ 743, 156
 \bibitem[]{} Masiero, J. R., Mainzer, A. K., Grav, T. et al. 2011, ApJ 741, 68
   \bibitem[]{} Maurette, M. 2005, in Micrometeorites and the mysteries of our origin. Springer-Verlag, Berlin/Heidelberg
  \bibitem[]{} Oey, J. 2006, MPBu, 33, 96
\bibitem []{} Perna, D., Dotto, E., Lazzarin, M. et al. 2010,  A\&A, 513, L4
   \bibitem[]{} Perozzi, E., Rossi, A., Valsecchi, G. B. 2001, P\&SS, 49, 3
   \bibitem[]{} Perozzi, E., Binzel, R. P., Rossi, A., Valsecchi, G. B. 2010, European Planetary Science Congress 2010, 750
  \bibitem[]{} Popescu, M., Birlan, M., Nedelcu, D. A. 2012, A\&A, 544, 130
   \bibitem[]{} Pravec, P., Harris, A.W.,  Kušnirák, P.,  Galád, A., Hornoch, K. 2012, Icarus, 221, 365
   \bibitem[]{} Prussing, J.E., 1992, JGCD, 15, 1037
   \bibitem[]{} Rayner, J. T., Toomey, D. W., Onaka, P. M. et al. 2003, PASP, 115, 362
   \bibitem[]{}    Reddy, V., Gaffey, M. J., Abell, P. A., Hardersen, P. S. 2006, 37th Lunar and Planetary Science Conference, LPI, 37, 1746
  \bibitem[]{} Reddy, V., Dyvig, R., Pravec, P. et al. 2007, 38th Lunar and Planetary Science Conference, LPI, 38, 1239
   \bibitem[]{} Sanchez, J. A., Reddy, V., Nathues, A. et al. 2012, Icarus, 220, 36 
   \bibitem[]{} Sanchez, J. A., Michelsen, R., Reddy, V., Nathues, A. 2013, Icarus 225, 131
  \bibitem[]{}         Shestopalov, D. I. \& Golubeva, L. F., 2011, 42nd Lunar and Planetary Science Conference,  LPI,42, 1608, 1028
   \bibitem[]{} Shoemaker, E. M. \& Helin, E. F., 1978, NASCP, 2053, 245
\bibitem[]{} Thomas, C. A., Trilling, D. E., Emery, J. P. et al. 2011, AJ, 142, 85
\bibitem[]{}Zimmer, A. K. \& Messerschmid, E. 2011, AA 69, 11, 1096



\end{thebibliography}
\end{document}